\documentclass[fleqn,usenatbib]{mnras}

\usepackage{newtxtext,newtxmath}
 
\usepackage[T1]{fontenc}
\usepackage{ae,aecompl}
\usepackage[normalem]{ulem}

\usepackage{graphicx}	
\usepackage{amsmath}	
\usepackage{amssymb}	
\usepackage{makecell}

\newcommand{\oiii}{[O{\sc iii}]}
\newcommand{\oii}{[O{\sc ii}]}

\title[An IR quasar lacking narrow optical emission lines]{A candidate Optically Quiescent Quasar lacking narrow emission lines}

\author[Greenwell et al.]{Claire Greenwell,$^{1}$\thanks{E-mail: c.l.greenwell@soton.ac.uk}
Poshak Gandhi,$^{1}$
Daniel Stern,$^{2}$
Peter Boorman,$^{3,1}$
Yoshiki Toba,$^{4,5,6}$
\newauthor
George Lansbury,$^{7}$
Vincenzo Mainieri,$^{7}$
Christopher Desira$^{8}$
\\
$^{1}$School of Physics \& Astronomy, University of Southampton, Highfield, Southampton SO17 1BJ, UK.\\
$^{2}$Jet Propulsion Laboratory, California Institute of Technology, 4800 Oak Grove Drive, Mail Stop 169-221, Pasadena, CA 91109, USA.\\
$^{3}$Astronomical Institute, Academy of Sciences, Boční II 1401, CZ-14131 Prague, Czechia.\\
$^{4}$Department of Astronomy, Kyoto University, Kitashirakawa-Oiwake-cho, Sakyo-ku, Kyoto 606-8502, Japan.\\
$^{5}$Academia Sinica Institute of Astronomy and Astrophysics, 11F of Astronomy-Mathematics Building, AS/NTU, No.1, Section 4, Roosevelt Road, Taipei 10617, Taiwan.\\
$^{6}$Research Center for Space and Cosmic Evolution, Ehime University, 2-5 Bunkyo-cho, Matsuyama, Ehime 790-8577, Japan.\\
$^{7}$European Southern Observatory, Karl-Schwarzschild-Strasse 2, D-85748
Garching, Germany.\\
$^{8}$Institute of Astronomy, University of Cambridge, Madingley Road, Cambridge CB3 0HA, UK.
}

\date{Accepted XXX. Received YYY; in original form ZZZ}

\pubyear{}

\begin{document}
\label{firstpage}
\pagerange{\pageref{firstpage}--\pageref{lastpage}}
\maketitle

\begin{abstract}
Many Active Galactic Nuclei (AGN) surveys rely on optical emission line signatures for robust source classification. There are, however, examples of luminous AGN candidates lacking such signatures, including those from the narrow line region, which are expected to be less susceptible to classical nuclear (torus) obscuration. Here, we seek to formalise this subpopulation of AGN with a prototypical candidate,  SDSS\,J075139.06+402810.9. This shows IR colours typical for AGN, an optical spectrum dominated by an early type galaxy continuum, an \oiii\,$\lambda$5007\AA\  limiting flux about two dex below Type\,2 quasars at similar IR power, and a $k$-corrected 12\,\micron\ quasar-like luminosity of $\sim$\,10$^{45}$ erg\,s$^{-1}$. These characteristics are not consistent with jet and host galaxy dilution. A potential scenario to explain this AGN quiescence in the optical is a sky-covering “cocoon” of obscuring material, such that the AGN ionising radiation is unable to escape and excite gas on kpc scales. Alternatively, we may be witnessing the short phase between recent triggering of obscured AGN activity and the subsequent narrow line excitation.
This prototype could define the base properties of an emerging candidate AGN subtype -- an intriguing  transitional phase in AGN and galaxy evolution.

\end{abstract}




\section{Introduction}

Understanding Active Galactic Nuclei (AGN) growth and evolution remains an active research theme. Much of the uncertainty and contention arises from the fact that AGN selection is a non-trivial problem. Broadband multiwavelength emission originates over a vast range of (logarithmic) physical scales of accreting and outflowing matter around the central engine, and is further sculpted by obscuration and scattering due to dust and gas. \citep[e.g. ][]{brandt_cosmic_2015, hickox_obscured_2018} -- AGN that are obvious in one selection regime are well hidden in others. If we are to find and understand the complete AGN population, broadband strategies are therefore necessary.

The orientation-based unification paradigm has helped enormously in providing structure to this quest \citep[e.g. ][]{antonucci_unified_1993, netzer_revisiting_2015, almeida_nuclear_2017}. However, it has also long been clear that this cannot be the whole story. In particular, AGN have been shown to grow together with their host galaxies, on average, and the influence of time-dependent changes to their environment remains ill-understood. Finding AGN at various evolutionary phases requires not only probing to higher redshifts, but also adapting and combining common AGN selection strategies in new directions. 

In certain models of AGN evolution, for instance, significant growth occurs while the central engine is completely enshrouded by obscuring material with $\sim$\,4$\pi$ sky covering factors  \citep[e.g.][]{fabian_obscured_1999}. Well-tested techniques based on searching for isotropically emitted AGN signatures would be ineffective to find such `cocooned' phases. These signatures include the narrow optical emission lines, in particular \oiii\,$\lambda$5007\AA. These forbidden lines arise from photoexcitation in the Narrow Line Region (NLR) on kpc scales. They arise above the classical unified torus, and are thought to be only modestly impacted by host galaxy reddening \citep[e.g., ][]{baldwin_classification_1981, risaliti_distribution_1999, zakamska_candidate_2003}. Dusty NLRs have been discussed for years (e.g., \citealt[][]{netzer93}), but cocooned AGN, in contrast, would show a complete absence of lines to deep limits if extreme covering factors can approach unity. Searches for such cocooned AGN could be promising at wavelengths less susceptible to dust \citep[e.g. ][]{gandhi_very_2002, ueda07, imanishi_spitzer_2010}, but the prevalence of such populations still remains unclear, especially at the luminous end.

Similarly, AGN undergoing transitions may fail to be recognised as such. A sudden onset of accretion activity on to a supermassive black hole would require time to manifest itself on kpc scales, due to the light-travel time to the NLR. During this phase, AGN may appear to be `optically elusive', and finding such objects can constrain AGN duty cycles \citep[e.g. ][]{schawinski_observational_2007}. Comprehensive searches of such populations of objects are thus vital to constrain AGN at important junctions in their evolution. 

As part of a systematic effort to identify and formalise such a population, we have begun a search for optically quiescent quasars (OQQ). Here, we present our first prototypical example, lay out the framework for our search, and discuss implications of finding more such objects. The full sample will be presented in a forthcoming work (Greenwell et al., in prep.). We assume a flat cosmology with $H_{\rm 0}$\,=\,67.4\,km\,s$^{-1}$\,Mpc$^{-1}$ and $\Omega_\Lambda$\,=\,0.685 \citep*{planck_collaboration_planck_2020}.

\section{Selection of a prototype}

With the ansatz of a cocooned AGN (an AGN enshrouded in obscuring matter; CAGN) in mind as a guide, we assume that typical optical and ultraviolet emission signatures will be reddened and scattered.
Optical spectroscopy would therefore indicate these objects to be similar to “normal” galaxies.  In order to efficiently block all lines-of-sight with the least mass of obscuring material, the size of any such cocoon must be small, plausibly similar to the canonical pc-scale torus. Many studies have indicated that reprocessed emission from these tori appears to be effectively optically-thin in the mid-infrared (MIR). Several physical models have been proposed to explain this, but the salient detail of relevance here is that the MIR is largely isotropic, irrespective of obscuring geometry \citep[e.g. ][ and references therein]{gandhi_resolving_2009}. If the same holds for CAGN, then MIR selection of OQQ is a promising route to uncovering them. 

We used infrared data from the \textit{Wide-Field Infrared Survey Explorer} (\textit{WISE}, \citealt{wright_wide-field_2010}), which carried out an all-sky survey in four bands, centred on wavelengths of 3.4, 4.6, 12, and 22\,\micron. Redshifts and optical classifications were obtained from the fifteenth data release of the Sloan Digital Sky Survey (SDSS DR15, \citealt{aguado_fifteenth_2019}). Targets with good quality spectra, not classified by SDSS as stars or quasars, and within a redshift range that would put the \oiii\ emission line within the spectrum ($z < 1.08$; observed wavelengths $\sim$ 5007\AA\ to 10400\AA) were cross matched with objects from the \textit{AllWISE} catalogue \citep{wright_wide-field_2010}. The resulting candidates included a large number of spectra that were optically classified as normal galaxies. MIR classification as AGN was carried out by using well-tested colour thresholds. Specifically, we chose to use the colour threshold $W1-W2 \geq 0.8$, shown by \citet{stern_mid-infrared_2012} to be an efficient selector of luminous AGN. 
To ensure that the AGN dominates above stellar emission and star formation, we required a rest frame $k$-corrected monochromatic 12\,\micron\ luminosity of $L_{12}\,\ge 3\,\times\,10^{44}$\,erg\,s$^{-1}$. This is typical of quasars with $L_{\rm Bol}$\,$\approx$\,10$^{45}$\,erg\,s$^{-1}$, and distinguishes our study from searches for so-called X-ray bright optically normal galaxies (XBONGs), which appear well explained by host galaxy dilution of less luminous AGN \citep[e.g. ][]{moran_hidden_2002}. 

The final requirement was the lack of any significant \oiii\ emission in the SDSS spectra. This procedure yielded several hundred candidates whose properties will be explored in an upcoming work (Greenwell et al., in prep.). Here, we present a first candidate, representative of our sample in terms of its high bolometric luminosity, reliability of redshift and optical spectral type, as well as a deep limit on the (expected) optical emission line flux. 

\begin{figure*}
    \includegraphics[width=0.85\textwidth]{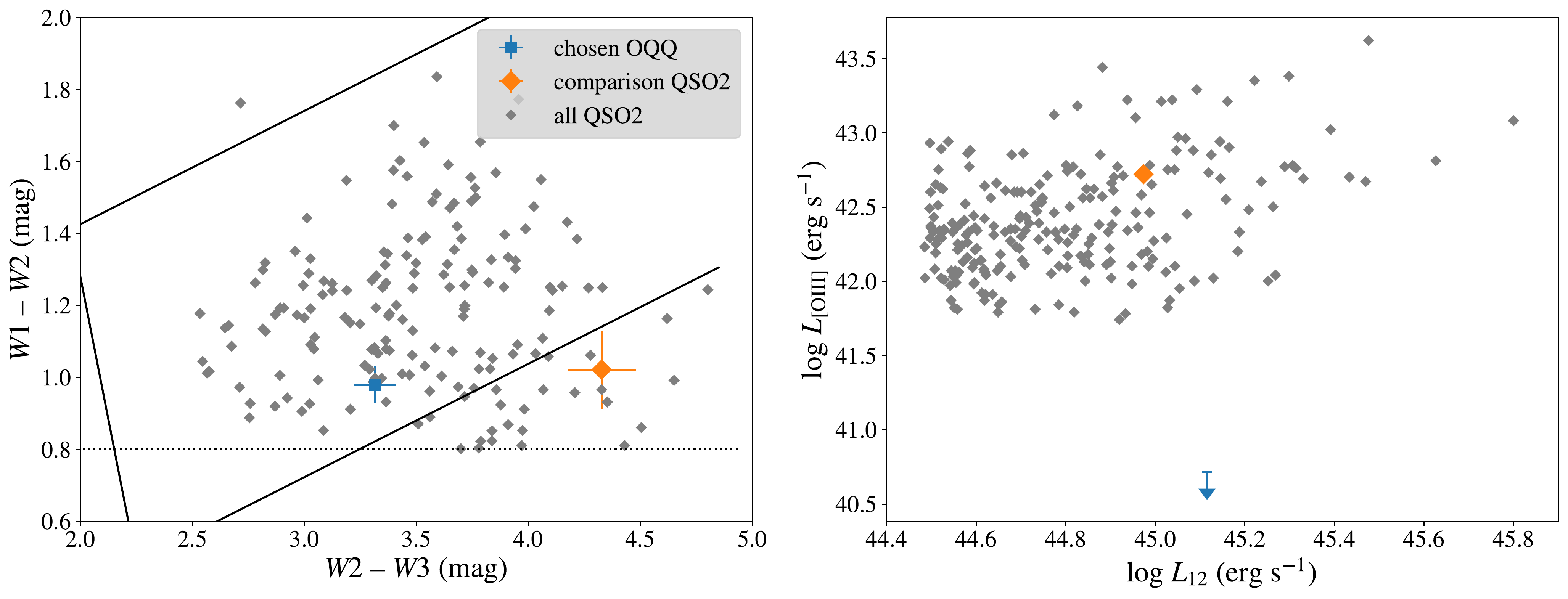}
    \caption{(left) The \textit{WISE} colours of the prototype OQQ plotted on the MIR colour-colour plane. The dotted line shows the Stern cut \citep{stern_mid-infrared_2012}; which all candidates pass by selection. The solid black lines show the AGN wedge proposed by \citet{mateos_using_2012}. Shown for comparison are QSO2s. (right) Upper limit of \oiii\ luminosity of OQQ (blue dash) compared to the measured \oiii\ fluxes of QSO2s.}
    \label{fig:wisecolours}
\end{figure*}

As a better understood `control' group with similar (but complementary) selection in terms of key parameters we choose to use the population of Type 2 quasars (QSO2s), as selected by \cite{reyes_space_2008}. These objects are bright obscured (classical) AGN, similar in redshift and broad-band luminosity range to our OQQ. The key exception is the presence, by design, of strong emission lines, particularly \oiii, in QSO2s.  

\section{Object Properties}

\begin{figure*}
	\includegraphics[width=0.65\textwidth, angle=270]{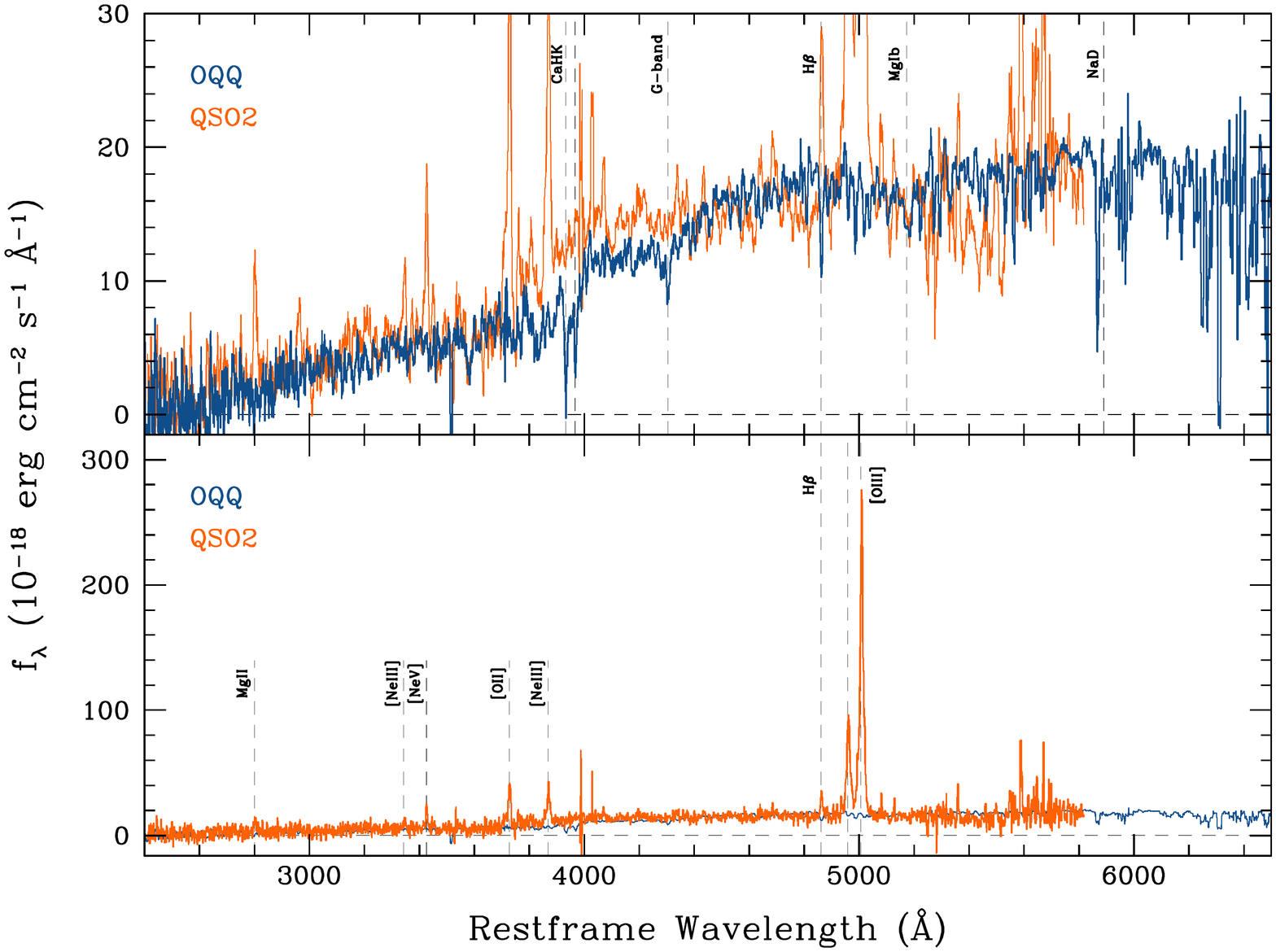}
    \caption{Spectra, obtained by SDSS, of the primary OQQ candidate, J0751, compared to the QSO2 J1256. The top panel is scaled to show the OQQ (blue), which is a typical early-type galaxy with stellar absorption features and a strong D4000 break. Note that the QSO2 (orange) shows a similar continuum shape, with stronger continuum redwards of \oii. The bottom panel is scaled to show the QSO2, which has multiple strong, narrow emission lines}
    \label{fig:spectracomp}
\end{figure*}

The prototype OQQ candidate chosen is SDSS J075139.06+402810.9 (hereafter J0751). We also choose a QSO2 close in redshift and luminosity to use as a direct comparison -- SDSS J125612.97+144121.0. Ideally we would also match the SFR of the two objects in order to minimise host galaxy dilution -- however due to the uncertainty in assessing the SFR of J0751 (see Section \ref{sec:agnfitter}), at this stage we focused only on matching the most important properties. Spectra of these two objects are presented in Figure \ref{fig:spectracomp}, showing the stark difference in \oiii\ strength between the two object types. Note also the similarity in continuum shape, with the main difference being emission line strength in the QSO2. Table \ref{tab:objdetails} presents basic information about the two objects. 

\begin{table}
	\centering
	\caption{Basic information about the primary OQQ candidate and the QSO2 closest in redshift and 12\,\micron\ luminosity. Column details: (1) object type, (2) object name as known in NED, (3) redshift, (4) rest frame 12 \micron\ luminosity.}
	\label{tab:objdetails}
	\begin{tabular}{ccrc}
\hline
  \thead{Type\\(1)}  &  \thead{NED name\\(2)}   &   \thead{$z$\\(3)} &  \thead{$\log L_{12} / {\rm{erg}}\, {\rm{s}}^{-1}$\\(4)}  \\
\hline
        OQQ         & SDSS J075139.06+402810.9 &              0.587 &                     45.12 $\pm$ 0.05                      \\
        QSO2         & SDSS J125612.97+144121.0 &              0.580 &                     45.04 $\pm$ 0.05                      \\
\hline
\end{tabular}
\end{table}

\begin{table}
	\centering
	\caption{Stellar mass and star formation rates of J0751 and J1256. Column details: (2) SFR derived from \oii\ luminosity \citep{kennicutt_star_1998}, (3), (4) SFR and stellar mass according to \texttt{agnfitter} (see Section \ref{sec:agnfitter}).}
	\label{tab:objsfr}
	\def\arraystretch{1.5}
	\begin{tabular}{cccc}
\hline
  \thead{Object\\(1)}  &  \thead{[OII] based SFR\\ $M_\odot$ per year\\(2)}  &  \thead{\texttt{agnfitter} SFR\\ $M_\odot$ per year\\(3)}  &  \thead{stellar mass\\ log $M_\odot$\\(4)}  \\
\hline
 \rule{0pt}{3ex} J0751 &                   <0.40$\pm$0.11                    &                 0.001$^{+0.216}_{-0.001}$                  &                    11.15$^{+0.11}_{-0.04}$                     \\
 \rule{0pt}{3ex} J1256 &                    5.11$\pm$1.46                    &                   1.69$^{+0.11}_{-0.14}$                   &                    10.74$^{+0.03}_{-0.04}$                     \\
\hline
\end{tabular}
\end{table}

\subsection{MIR Colour and Luminosity}

\begin{figure*}
	\includegraphics[width=0.84\textwidth]{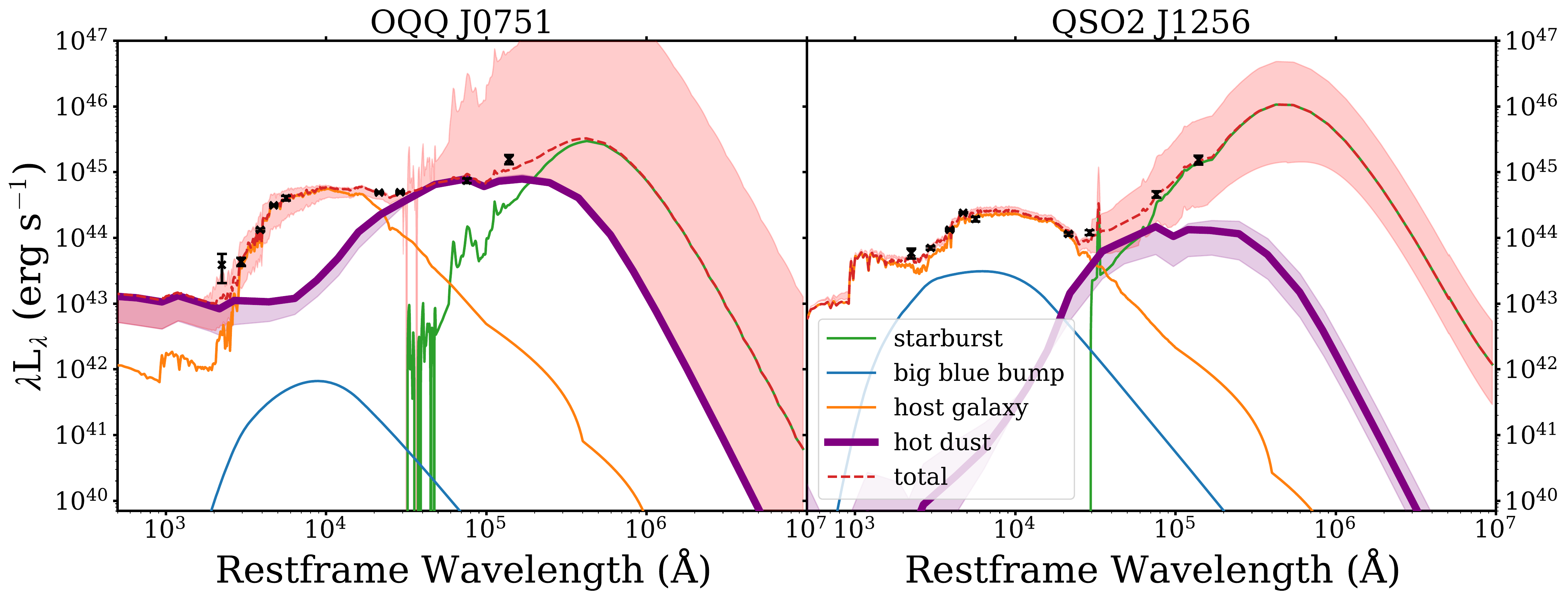}
    \caption{\texttt{agnfitter} results. Known data is shown as black crosses, and the lines show contributions from starburst (green), host galaxy (yellow), hot dust emission (purple), and big blue bump (blue; originates from the accretion disc). AGN-heated dust emission is highlighted in dark purple. The 1 sigma uncertainty range is also shown for the hot dust and total only. Blue emission directly from the accretion disc is much more prominent for J1256 than for J0751, whereas the hot dust emission from J0751 is required to be higher.}
    \label{fig:sed_fit}
\end{figure*}

The most convincing evidence for the presence of an AGN lies in the MIR. A k-corrected 12\,\micron\ luminosity of $\sim$ 10$^{45}$ erg s$^{-1}$ (estimated by linear interpolation between \textit{WISE} \textit{W3} and \textit{W4} measurements) implies a very low chance that this object lacks an AGN. SED fitting (see Section \ref{sec:agnfitter} and Figure \ref{fig:sed_fit}) shows that a significant contribution from AGN-heated dust is required to reproduce the observed MIR data. The MIR colour selection criterion has a predicted reliability of 95\% \citep{stern_mid-infrared_2012}, depending on the reliability of the source data for this. There is a small chance that similar colour and luminosity may be caused by star formation. If this were the case we would expect to see bright emission lines typically associated with star formation, such as \oii\, as the distribution of star formation throughout the galaxy (rather than localised at the nucleus) means that they are unlikely to be blocked by the same obscuring matter. Conversely, if the star formation is localised at the nucleus we may not expect to detect strong \oii\ emission, but the SED fitting (see Section \ref{sec:agnfitter} should provide a better estimate of SFR. We also see very little radio emission (an upper limit on radio flux is available from FIRST \citep{becker_first_1995}: <0.695 mJy/beam).

\subsection{Spectral Energy Distribution (SED)}\label{sec:agnfitter}

We use \texttt{agnfitter} \citep{calistro_rivera_agnfitter_2016} to fit the SED of J0751 and determine the relative galaxy and AGN system contributions to the observed emission. This method uses the multiwavelength data available to find the most likely weighted contributions to the overall SED, from sources near the AGN and in the host galaxy. Lack of IR data at wavelengths redward of \textit{WISE} produces some uncertainty on these results, but we can use them to constrain further some properties of the target. The most important parameter for our analysis is the contribution to the SED from the AGN obscuring material in the MIR -- a significant amount of emission is needed from this component in order to reproduce the total. After experimenting with different MCMC lengths, we selected one sufficient to achieve an auto-correlation time indicating convergence for the OQQ AGN-heated dust component (see \citet{foreman-mackey_emcee_2013} \& \citet{calistro_rivera_agnfitter_2016} for details).

\texttt{agnfitter} makes two separate estimates of SFR -- one in the optical, and one in the FIR. The FIR SFR would provide some estimate of potential obscured star formation, but due to the lack of FIR data available for our fits, we discard this version and focus only on the optical. This estimate (and that of stellar mass) is obtained from host galaxy template parameters (for more detail, see \citealt{calistro_rivera_agnfitter_2016}). As shown in Table~\ref{tab:objsfr} the SFR obtained for J0751 is low, essentially an upper limit only, and is significantly higher for J1256. The uncertainty on the optical SFR is higher than for other \texttt{agnfitter} derived properties (excluding the FIR SFR) -- e.g. 6--8\% for J1256.

It is possible that the star formation may be located very close to the nucleus, and as such may have \oii\ obscured. If this were the case, a large SFR would still be necessary to explain the \textit{WISE} data without a significant AGN dust component.

\vspace*{-0.5cm}
\section{Context}

\subsection{Cocooned AGN growth and the merger paradigm}

AGN grow in proportion with the bulge of their host galaxies, on average. One possible evolutionary track that could explain this is merger-driven evolution \citep{fabian_obscured_1999, kormendy_coevolution_2013, hopkins_unified_2006}, where galactic mergers funnel large amounts of material inwards, boosting both AGN growth and star formation \citep[e.g.][]{, springel_modelling_2005}. 
 
The influx of material provides the fuel for AGN as well as star formation, and also obscures the nucleus. As the AGN brightens, radiation pressure clears away the obscuring material, revealing the nucleus in a bright unobscured phase. Eventually, when all the gaseous fuel is exhausted, the AGN returns to its dormant state. One plausible interpretation of the OQQ characteristics is a short phase in this cycle, when the infalling material has ignited the AGN, but it still remains obscured along all lines-of-sight. If this scenario were appropriate for J0751, we might expect to see (a) signs of recent merger activity, and (b) significant ongoing star formation, as seen in some local samples including starburst-AGN composites \citep{goulding09}. 

Optical images from SDSS and \textit{Pan-STARRS} show a compact, red galaxy, with no apparent ongoing merging activity. Higher resolution images may be able to shed more light on the recent history of this galaxy -- \textit{Pan-STARRS} provides a median image quality of just above 1 arcsecond \citep{magnier_pan-starrs_2020} ($\sim$11\,kpc). The red colour, presence of an early-type galaxy spectral continuum, and lack of optical emission lines in J0751 (Figure \ref{fig:spectracomp}) imply that there is very little star formation. The prominent 4000\,\AA\ break also distinguishes J0751 from other featureless systems such as BL Lacs, where continuum emission is dominated by nuclear jet emission. 

Based on the broadband photometry rather than emission line strengths, SED modelling by {\tt agnfitter} gives slightly different values of $0.001$ and 1.69 $M_\odot$ per year, respectively. In principle H$\alpha$ could also provide useful constraints here, but is shifted redward of our current spectral coverage. High quality near-IR spectroscopy (to determine H$\alpha$) and longer wavelength photometric measurements (to constrain the \texttt{agnfitter} SED decomposition) would provide a more accurate value for the SFRs and further constrain the presence of an AGN.

The lack of star formation does not easily fit into the merger-driven AGN and host galaxy evolutionary  paradigm. If this is the result of obscuration, it would require that any ongoing star-formation (SF) is also obscured, or that there be a delay between AGN ignition and subsequent SF triggering (or vice versa). Neither of these solutions is likely. There is no reason to expect SF triggering to await AGN triggering if there is plentiful gas supply in the environment. 

\vspace*{-0.5cm}
\subsection{Recently triggered AGN activity?}
A second possible origin that can explain our observation is the young AGN theory -- if the “switch on” occurs through secular processes local to the AGN, and is recent enough that the NLR (the usual origin of the narrow \oiii\ emission) is not yet active. This would not require the presence of a cocoon, but instead line-of-sight obscuration of the accretion disc and the broad line region. 

Typical AGN feeding timescales are thought to be of order $\sim$\,10$^5$  years \citep{schawinski_active_2015}. If we happen to have detected the AGN within a few years of its trigger, radiation would have had time to be reprocessed in the pc-scale torus, but not yet excite the kpc-scale NLR, through neither radiative nor collisional processes. In many ways, this would be the opposite of the preferred scenario to explain Voorwerp systems where strong, offset NLR emission is seen with no strong nuclear activity evident \citep[e.g.][]{sartori_joint_2018}.  The Voorwerp are believed to be AGN that have recently turned off, but the NLR is sufficiently distant to be responding to earlier AGN activity. There are also suggestions that some of these systems contain hidden AGN - \citet{lansbury_storm_2018} show a Voorwerp system where the central AGN is luminous but hidden. Typical recombination times in nearby QSO2s are estimated to be between a few years to a few hundred years, depending upon gas density  \citet{trindade_falcao_hubble_2020}, so a steady source of excitation is important. A final possibility is that the NLR gas is no longer confined and has dissipated or over-ionised over time. Though this cannot be excluded, it is unclear why this should be the case in these particular objects. The AGN would still need to be obscured along the line-of-sight in order to extinguish BLR and disc continuum signatures. 

We have presented the first prototype in an effort to formalise the population of OQQs. Similar objects have cropped up in various other studies, but have not yet been systematically collated. For instance, \citet{hviding_characterizing_2018} select AGN using WISE, but aim for a different part of the WISE colour-colour plane that restricts their targets to the most heavily obscured objects. There is no overlap, however some of this sample may be interesting cousins to OQQs. Most (70\%) of their targets are actually identified as AGN through common optical emission line diagnostics - all but one of these have detectable \oiii\ lines. \citet{glikman_first-2mass_2012} use cross-matches between FIRST \citep{becker_first_1995} and 2MASS \citep{skrutskie_two_2006}. \citet{ueda07} selected `geometrically thick torus' AGN by looking for low-scattering X-ray fractions in the Swift/BAT survey; these could be lower luminosity cousins to OQQs. All such objects, although not identical to OQQs, could represent a similar population, or a close phase in a lifecycle containing both, so a population wide study is crucial and will be the subject of a follow-up work.

There remains a possibility that J0751 does not contain an AGN, but is a coincidence of emission from other sources that give the impression of one. However, to reproduce the emission we have without an AGN, the spectrum would need to be heavily dominated by starburst activity in the IR. The weakness of the non-\oiii\ emission lines present make this extremely unlikely.

Polarimetric observations, infrared spectroscopy and X-ray observations could help to reveal the true nature of these objects. X-ray observations would be particularly useful to prove the presence of an accreting nucleus; if the line-of-sight obscuring column is Compton-thick then hard X-ray data, such as from NuSTAR, would be necessary.

Irrespective of the true nature of these objects, they appear to represent an ill-studied phase of AGN (or AGN-like) activity. It should eventually be possible to constrain the duty cycle of this phase through demography of this population. 

\vspace{-6mm}

\section*{Acknowledgements}

We thank the referee for their helpful comments and suggestions. CG is supported by a University of Southampton Mayflower studentship. PG acknowledges support from STFC and a UGC-UKIERI Thematic partnership. P.B. acknowledges financial support from the STFC and the Czech Science Foundation project No. 19-05599Y. This publication makes use of data products from the {\it Wide-field Infrared Survey Explorer}, which is a joint project of the University of California, Los Angeles, and the Jet Propulsion Laboratory/California Institute of Technology, funded by the National Aeronautics and Space Administration.

\vspace{-6mm}

\section*{Data Availability Statement}

The data underlying this article are publicly available from the \textit{WISE} All-Sky Survey and SDSS DR15.

\vspace{-6mm}


\bibliographystyle{mnras}
\bibliography{lib.bib} 




\appendix


\bsp	
\label{lastpage}
\end{document}